\documentclass[10pt,final]{siamltex}
\usepackage{amsmath}
\newcommand{\diff}{\mathrm{d}}
\newcommand{\modelindent}{\hspace{2.2em}}
\makeatletter
\newcommand{\flushsection}[1]{%
  \par\refstepcounter{section}\setcounter{subsection}{0}%
  \addcontentsline{toc}{section}{\protect\numberline{\thesection}#1}%
  \vspace{1.1\baselineskip}%
  \noindent{\bfseries \thesection.\ #1.}\par\nobreak\vspace{0.7\baselineskip}%
}
\makeatother

\usepackage{bm}
\usepackage{amssymb,version}
\usepackage{cases}
\usepackage{color}
\usepackage{verbatim}
\usepackage{fancyvrb}
\usepackage{graphics}
\usepackage{epsfig}
\usepackage{epstopdf}
\usepackage{enumerate}
\usepackage{lipsum}
\usepackage{multirow}
\usepackage{array}
\usepackage{booktabs}
\usepackage{placeins}
\usepackage{float}
\usepackage{mathrsfs}

\usepackage{graphicx}
\usepackage{subfig}
\usepackage{svg}
\usepackage[colorlinks=true, allcolors=blue]{hyperref}
\usepackage{cleveref}

\allowdisplaybreaks

\newcommand{\normmm}[1]{{\left\vert\kern-0.25ex\left\vert\kern-0.25ex\left\vert #1
    \right\vert\kern-0.25ex\right\vert\kern-0.25ex\right\vert}}

        \renewcommand{\thethm}{\thesection.\arabic{thm}}

\begin{document}

     \title{The PICNN-Assisted Physics-Preserving Scheme for Thermodynamically Consistent Two-Phase Flow in Porous Media
\thanks{ This work is supported by the China Postdoctoral Science Foundation (grant 2025M783124) and Postdoctoral Innovation Program of Shandong Province (grant SDCX-ZG-202602017).} }

\author{
Yuanshuo Kong, Xue Wang$^{*}$, and Yujing Yan
\thanks{
School of Mathematics, Shandong University, Jinan, Shandong, 250100, P.R. China.}}
\thanks{Corresponding author. Email: wangxsdu@mail.sdu.edu.cn.
}
\graphicspath{{figures/},{../MAC-CS/figures/},}
\maketitle

\begin{abstract}
In this paper, we develop a physics-informed convolutional neural network (PICNN) assisted physics-preserving method for a thermodynamically consistent model of incompressible and immiscible two-phase flow in porous media. Following the physics-preserving prediction-correction scheme of Li et al. \cite{li2025class}, the prediction step is performed by a PICNN trained with finite-volume residuals, where the interfacial fluxes are evaluated by the two-point flux approximation (TPFA) using two-point difference quotients of neighboring cell-centered unknowns to approximate interfacial normal gradients. The PICNN output is further corrected by a post-processing procedure to obtain energy-stable, mass-conservative, and bounds-preserving solutions. Numerical results show that the finite-volume residuals trained PICNN can replace the traditional prediction solver within the physics-preserving framework. Compared with conventional physics-informed neural networks (PINNs), the PICNN better captures local spatial interactions between each control volume and its neighboring cells, while the finite-volume residuals accommodate discontinuous permeability fields and interfacial flux continuity.

\end{abstract}


\flushsection{Introduction}
\noindent


The modeling and numerical simulation of incompressible and immiscible two-phase flow in porous media are important in reservoir engineering and groundwater transport \cite{joekar2012analysis,helmig1997multiphase}. Classical computational methods for multiphase flow in porous media include finite-volume, mixed-dimensional, and multiscale formulations \cite{chen2006computational,reichenberger2006mixed,tomin2013hybrid}. In particular, finite-volume methods are widely used because their control-volume formulation is naturally conservative and is well suited to discontinuous material coefficients \cite{eymard2000finite}.

In recent years, deep learning has provided flexible approximation tools for high-dimensional maps and data-driven modeling \cite{lecun2015deep,goodfellow2016deep}. Deep learning based partial differential equation (PDE) solvers and operator learning methods have also attracted increasing interest \cite{sirignano2018dgm,e2018deep}; see also the Fourier neural operator approach \cite{li2021fourier}. Its application to the numerical solution of partial differential equations (PDEs) has been extensively studied in physics-informed learning \cite{raissi2019physics,karniadakis2021physics}, together with recent reviews and numerical analysis \cite{de2024PINNsurvey,luo2025PINNreview}. Applications in geotechnical and geoengineering simulations have also been reported \cite{chen2024PINNgeo}. Specifically, the PINN proposed by Raissi et al. \cite{raissi2019physics} incorporates the residuals of governing equations and the initial and boundary conditions into the loss function, providing a new framework for solving forward and inverse problems governed by PDEs. Compared with purely data-driven neural networks \cite{zhu2018bayesian,li2021fourier}, PINN introduces physical information into the training process and can work with limited labeled data. However, for strongly discontinuous permeability fields, traditional PINNs based on pointwise automatic-differentiation residuals do not explicitly represent conservative flux balance across interfaces. Therefore, the loss function is constructed from finite-volume discrete residuals rather than pointwise continuous PDE residuals.

 Apart from the residual formulation, the network architecture should also exploit the grid structure of porous-media flow variables. In two-phase porous-media flow, pressure, saturation, permeability, and porosity are usually represented as grid-based fields, which makes CNNs suitable for extracting local spatial features from such variables. Related deep-learning surrogate and operator-learning methods have been developed for parametric flow problems and uncertainty quantification \cite{zhu2018bayesian,wang2021mlcnnporous}, inverse problems \cite{mo2019deep}, and related PDE models \cite{li2021fourier}.
More recently, Zhang et al. \cite{zhang2023physics} proposed PICNN for the simulation and prediction of two-phase Darcy flow in heterogeneous porous media, where FVM is used to approximate the governing equation residual in the loss function, so that the CNN can be trained as a physics-informed predictor for the pressure field. 
Related PICNN methods have also been developed for porous media flow problems with time-varying controls \cite{chen2024transfer}. The key advantage of PICNN over standard PINN is that it encodes conservative flux balances across cell interfaces through FVM residuals, and it avoids the computational cost and potential instability of automatic differentiation for complex physical coefficients. Nevertheless, these works mainly aim to construct neural network solvers or surrogate models, while structure-preserving properties in thermodynamically consistent models, such as energy dissipation, mass conservation correction, and saturation bounds preservation, are not the main focus.

Thermodynamically consistent two-phase flow models based on free energy have attracted increasing attention \cite{gao2020thermodynamically,abels2012thermodynamically}.
Such models usually use free energy to characterize interfacial interactions or capillary effects, and they can satisfy the energy dissipation law at the continuous level. Therefore, when these models are numerically discretized, it is necessary not only to ensure the accuracy of the numerical solution, but also to preserve physical properties as much as possible. To achieve these goals, structure-preserving numerical methods have been developed for thermodynamically consistent two-phase flow models \cite{kou2022energy,kou2023energy,li2025class}. 
Energy-stable linearization techniques such as invariant energy quadratization have been developed for phase-field models and thermodynamically consistent flow systems \cite{yang2017numerical,kou2022energy}. For incompressible two-phase flow, mass-conservative and physics-preserving IMPES schemes have also been constructed \cite{chen2019fully,chen2021new}, and related multiscale extensions have been proposed for highly heterogeneous media \cite{wang2025physics}.
To further improve the ability of numerical schemes to preserve physical properties, scalar auxiliary variable (SAV) approaches have been developed for gradient systems and related dissipative models \cite{shen2018sav,huang2020highly}.
Generalized SAV approaches further extend this idea to broader gradient systems \cite{cheng2021generalized}.
Based on these ideas, Li et al. \cite{li2025class} combined the modified generalized SAV (mGSAV) method with the Lagrange multiplier (LM) method, and used the Karush--Kuhn--Tucker (KKT) conditions to handle the energy dissipation law, mass conservation law, and boundedness of saturations. The constructed physics-preserving numerical schemes only need to solve one linear system and a nonlinear algebraic equation with negligible
computational cost at each time step. 
This correction strategy is related to bound-preserving LM techniques \cite{cheng2022lagrange} and to finite-volume schemes for related phase-field or gradient-flow models that combine bound preservation with discrete energy dissipation \cite{bailo2023unconditional}. It is also connected in spirit to classical maximum-principle-preserving and positivity-preserving methods for nonlinear conservation laws and fluid models \cite{zhang2010maximum,zhang2010positivity}.
However, from the viewpoint of computational implementation, this type of method still mainly relies on traditional numerical solvers to complete the prediction step. When the computational grid is large or many groups of parameter simulations are required, the computational cost remains a concern.

Motivated by physics-informed learning and structure-preserving numerical schemes, this work develops a PICNN-based prediction--correction framework for incompressible and immiscible two-phase flow in porous media. The governing equations are discretized using a finite-volume method, where pressure and wetting saturation are treated as the primary unknowns. At each time step, the PICNN maps the previous state, including pressure, saturation, spatially distributed permeability and porosity fields, and normalized physical coordinates, to an intermediate prediction of pressure and saturation at the next time level. The permeability and porosity are treated as prescribed spatial fields. This prediction is then corrected by enforcing discrete mass conservation, energy stability, and saturation bound constraints to ensure physical consistency.

The proposed method can be combined with transfer learning strategies during the training of the CNN at each time step to further improve computational efficiency, as demonstrated in PICNN-based porous-media simulations and related physics-informed learning studies \cite{chen2024transfer,guo2025CTL_PINN}. Overall, the proposed method integrates data-driven prediction with physically constrained correction to achieve a balance between computational efficiency and numerical stability in two-phase flow simulations.

The main work of this paper includes the following four aspects:
\begin{itemize}
    \item We construct a PICNN prediction module with pressure and wetting  saturation as joint outputs. Compared with fully connected neural networks, CNNs use local connectivity and weight sharing, and are therefore more suitable for large grid-based two-phase flow fields.
    \item The PICNN is trained by finite-volume residuals constructed from the conservative pressure and saturation balance equations, where the interfacial fluxes are evaluated by TPFA and two-point difference quotients.
    \item Mass conservation, energy stability, and boundness-preserving of two phase saturations corrections are added after the PICNN prediction, forming a hybrid solution framework.
    \item Numerical examples are presented to examine the saturation distribution, mass conservation error, energy behavior, and bounds-preserving property of the corrected solutions, thereby evaluating the effectiveness and limitations of the proposed method.
\end{itemize}

The remainder of this paper is organized as follows. In Section 2, we introduce the thermodynamically consistent model for incompressible and immiscible two-phase flow in porous media, including its free-energy structure, mass conservation law, saturation bounds, and energy dissipation law, and then outline the PICNN framework. In Section 3, we present the finite-volume residual training of the PICNN predictor, the subsequent physics-preserving correction steps, and the resulting structure-preserving properties. In Section 4, numerical experiments are provided to validate the accuracy, bounds-preserving property, energy behavior, and mass conservation of the proposed method. Finally, concluding remarks are given in Section 5.
Throughout this paper, $(\cdot,\cdot)$ denotes the $L^2$ inner product over $\Omega$ for scalar, vector, or matrix functions, and $\langle\cdot,\cdot\rangle$ denotes the $L^2$ inner product on the boundary $\partial\Omega$. The notation $\|\cdot\|$ stands for the $L^2$ norm.

\flushsection{Mathematical Model and Physical Structures}\label{sec-2}
Firstly, in Section 2.1, we present the thermodynamically consistent model for incompressible and immiscible two-phase flow in porous media, together with a reformulation of the governing equations and the associated free energy function to reduce the number of unknown variables. Then, Section 2.2 introduces a PICNN to approximate the simplified coupled governing equations.

\makeatletter
\let\CSMACsavedsubsection\subsection
\renewcommand\subsection{\@startsection{subsection}{2}{0pt}%
                                     {1.3ex\@plus .5ex \@minus .2ex}%
                                     {-.5em \@plus -.1em}%
                                     {\reset@font\normalsize\bfseries}}
\makeatother

\subsection{Two-phase flow model and energy dissipation}\leavevmode\par\nobreak\smallskip
Let us present the thermodynamically consistent two phase model consisting of the phase-wise mass balance equations \eqref{eq:model_mass}, Darcy law \eqref{eq:model_darcy}, and the algebraic saturation constraint \eqref{eq:model_constraint} as follows,
\begin{flalign}
&\modelindent\phi\frac{\partial S_\alpha}{\partial t}+\nabla\cdot\mathbf{u}_\alpha=q_\alpha,\quad \alpha=w,n, &&
    \label{eq:model_mass}\\[0.25em]
&\modelindent\mathbf{u}_\alpha=-\lambda_\alpha K\nabla(p+\mu_\alpha),\quad \alpha=w,n, &&
    \label{eq:model_darcy}\\[0.25em]
&\modelindent S_w+S_n=1. &&
    \label{eq:model_constraint}
\end{flalign}
Here $\phi$ is the porosity, the subscripts $w$ and $n$ represent the wetting and non-wetting phases, respectively. For each phase $\alpha$, $S_\alpha$ and $\mathbf{u}_\alpha$ denote its saturation and Darcy flux. $p$ denotes the global pressure, and the admissible saturation interval is prescribed by
\begin{flalign}
&\modelindent S_{rw}\leq S_w\leq 1-S_{rn},\quad S_{rn}\leq S_n\leq 1-S_{rw}, &&
    \label{eq:saturation_bounds}
\end{flalign}
where $S_{rw}$ and $S_{rn}$ are the residual saturations of the wetting and non-wetting phases. The permeability tensor $K$ is assumed to be symmetric and positive. In the subsequent discussion we take the isotropic case $K=K_0I$, with $K_0>0$ and $I$ the identity tensor. The mobility of phase $\alpha$ is given by
$\lambda_\alpha(S_\alpha)=\dfrac{k_{r\alpha}(S_\alpha)}{\eta_\alpha}>0,\, \alpha=w,n, $
where $k_{r\alpha}$ is the relative permeability, $\eta_\alpha$ is the viscosity, and $q_\alpha$ denotes the injection or production rate of phase $\alpha$.

Following the idea \cite{li2025class}, it is convenient to eliminate one saturation variable for the numerical approximation. Without loss of generality, we get rid of $S_n$ using the constraint \eqref{eq:model_constraint}. Let us introduce the phase pressure
\begin{flalign*}
&\modelindent p_\alpha=p+\mu_\alpha,\qquad \alpha=w,n. && 
\end{flalign*}
Then the capillary pressure is given by
\begin{flalign*}
&\modelindent p_c=\mu_n(S_w,S_n)-\mu_w(S_w,S_n)=p_n-p_w. &&
\end{flalign*}
Using the constraint \eqref{eq:model_constraint}, we define the free energy $F$ only dependent on $S_w$ as
\begin{flalign}
&\modelindent\begin{aligned}
F(S_w)=&\sigma_wS_w(\ln(S_w)-1)
       +\sigma_n(1-S_w)(\ln(1-S_w)-1)+\sigma_{wn}S_w(1-S_w),
\end{aligned} &&
    \label{eq:free_energy_reduced}
\end{flalign}
where $\sigma_w, \sigma_n>0$ and $\sigma_{wn}$ are energy coefficients.
The chemical potentials are defined as partial derivatives of the free energy $F$ with respect to the phase saturations. Thus, we just define the chemical potential with respect to the wetting-phase saturation as
\begin{flalign*}
&\modelindent\begin{aligned}
\mu(S_w)
=\frac{\partial F(S_w)}{\partial S_w}
&=\sigma_w\ln(S_w)-\sigma_n\ln(1-S_w)+\sigma_{wn}(1-2S_w)\\
&=\mu_w(S_w,S_n)-\mu_n(S_w,S_n)=p_w-p_n=-p_c .
\end{aligned} &&
\end{flalign*}
Consequently, the non-wetting phase pressure $p_n$ is
$p_n=p_w-\mu(S_w). $
Then, the system \eqref{eq:model_mass}-\eqref{eq:model_constraint} can be rewritten as the following reduced pressure-saturation formulation:
\begin{flalign}
&\modelindent\left\{
\begin{aligned}
-\nabla\cdot\left(\lambda_t(S_w)K\nabla p_w\right)
 +\nabla\cdot\left(\lambda_n(S_w)K\nabla\mu(S_w)\right)
&=q_t,\\
\phi\frac{\partial S_w}{\partial t}
 -\nabla\cdot\left(\lambda_w(S_w)K\nabla p_w\right)
&=q_w,
\end{aligned}
\right. &&
\label{eq:reduced_system}
\end{flalign}
where $\lambda_t(S_w)=\lambda_w(S_w)+\lambda_n(S_w)$ denotes the total mobility and $q_t=q_w+q_n$. Since $\lambda_w,\lambda_n>0$, one obtains $\lambda_t(S_w)>0$ for $S_w\in(0,1)$. 

The system \eqref{eq:reduced_system} satisfies the following energy stable equation 
\begin{flalign}
&\modelindent\frac{\diff\mathcal E(t)}{\diff t}
=\sum_{\alpha=w,n}\int_\Omega p_\alpha q_\alpha\,\diff x
 -\sum_{\alpha=w,n}\int_{\partial\Omega}p_\alpha\mathbf{u}_\alpha\cdot\mathbf{n}\,\diff s
 -\sum_{\alpha=w,n}\left\|(\lambda_\alpha K)^{1/2}\nabla p_\alpha\right\|^2, &&\label{eq:energy_identity_reduced}
\end{flalign}
where the total free energy is defined as
\begin{flalign}
&\modelindent \mathcal E(t)=\int_\Omega \phi F(S_w)\,\diff x. &&
\label{eq:total_energy_reduced}
\end{flalign}
In particular, for a closed system with $q_\alpha=0$ and $\mathbf{u}_\alpha\cdot\mathbf{n}|_{\partial\Omega}=0$, the following energy dissipation holds
\begin{flalign}
&\modelindent\frac{\diff\mathcal E(t)}{\diff t}
=-\sum_{\alpha=w,n}\left\|(\lambda_\alpha K)^{1/2}\nabla p_\alpha\right\|^2\leq 0 . &&
\label{eq:energy_dissipation_reduced}
\end{flalign}

\begin{figure}[H]
\centering
\includegraphics[width=0.82\textwidth]{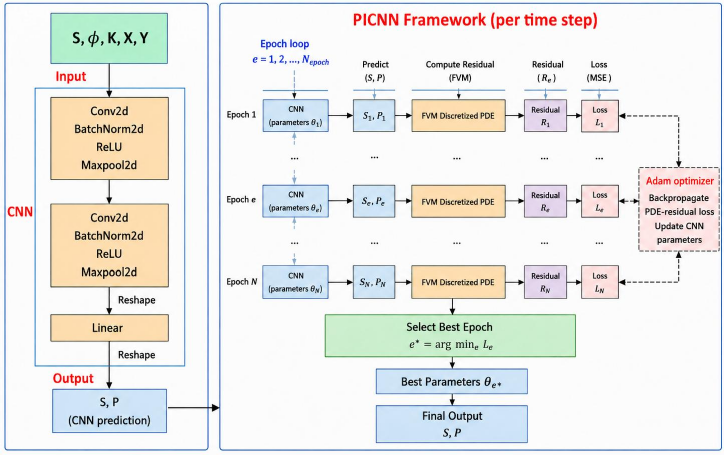}
\caption{Schematic illustration of the physics-informed CNN framework.}
\label{fig:picnn_framework}
\end{figure}

\subsection{Physics-Informed Convolutional Neural Network}\leavevmode\par\nobreak\smallskip

The CNN is used here as a grid-based approximation to the discrete-time evolution operator of the two-phase flow system. Given the physical fields at the current time level, including saturation $S$, pressure $P$, porosity $\phi$, permeability $K$, and the coordinate fields $(X,Y)$, the network produces the next-state prediction of saturation and pressure. In this sense, the CNN represents the nonlinear map induced by the discretized pressure--saturation equations rather than an isolated pointwise regression rule.

The use of convolutional layers is natural for porous-media flow variables because the unknowns and coefficients are stored on spatial grids and the dominant interactions are local flux exchanges across neighboring control volumes. A purely data-driven CNN, however, does not by itself encode the conservative flux balance, the pressure-saturation coupling, or the dependence of the phase motion on heterogeneous medium properties. These missing structures may lead to nonphysical deviations when the learned map is repeatedly applied in time.

To reduce this discrepancy, the CNN is embedded in a physics-informed framework. As illustrated in Fig.~\ref{fig:picnn_framework}, the network prediction at a training epoch, denoted by $(S_e,P_e)$, is substituted into the finite-volume discretization of the governing equations. The resulting residual measures the mismatch between the predicted fields and the discrete pressure-saturation balance. The network parameters are then updated by minimizing the residual-based loss through backpropagation with the Adam optimizer \cite{kingma2015adam}.

During training, the residual loss decreases over epochs, and the parameter set associated with the minimum loss is selected as $\theta^*$. The corresponding CNN output provides the intermediate pressure and saturation fields used by the correction procedure developed in Section 3. Thus, the PICNN couples the expressive capacity of convolutional operator learning with finite-volume residual constraints, so that the prediction remains connected to the conservative structure of the discretized two-phase flow model.

In the time-marching computation, transfer learning is used to reduce the training effort after the initial step. More precisely, the CNN is trained from an initial parameter set at the first time step, and the obtained parameter set $\theta^*$ is then used as the initial guess for the network training at the next time step. Since two neighboring time levels usually have similar pressure and saturation fields, the later-step networks can be fine-tuned with fewer epochs. In the numerical experiments, the first/later-step epoch settings are 500/100 epochs in Example 1 and 2000/100--200 epochs in Example 2.
\let\subsection\CSMACsavedsubsection
\makeatletter
\let\CSMACsavedsection\section
\let\CSMACsavedparagraph\paragraph
\renewcommand\section{\@startsection{section}{1}{0pt}%
                                     {1.1\baselineskip}%
                                     {0.7\baselineskip}%
                                     {\reset@font\large\bfseries}}
\renewcommand\subsection{\@startsection{subsection}{2}{0pt}%
                                     {1.3ex\@plus .5ex \@minus .2ex}%
                                     {0.8ex\@plus .2ex}%
                                     {\reset@font\normalsize\bfseries}}
\renewcommand\paragraph{\@startsection{paragraph}{4}{\parindent}%
                                     {1.6ex\@plus .5ex \@minus .2ex}%
                                     {0.75ex\@plus .2ex}%
                                     {\reset@font\normalsize\bfseries}}
\makeatother
\newskip\CSMACsavedabovedisplayskip
\newskip\CSMACsavedbelowdisplayskip
\newskip\CSMACsavedabovedisplayshortskip
\newskip\CSMACsavedbelowdisplayshortskip
\newdimen\CSMACsavedjot
\CSMACsavedabovedisplayskip=\abovedisplayskip
\CSMACsavedbelowdisplayskip=\belowdisplayskip
\CSMACsavedabovedisplayshortskip=\abovedisplayshortskip
\CSMACsavedbelowdisplayshortskip=\belowdisplayshortskip
\CSMACsavedjot=\jot
\setlength{\abovedisplayskip}{8pt plus 2pt minus 1pt}
\setlength{\belowdisplayskip}{8pt plus 2pt minus 1pt}
\setlength{\abovedisplayshortskip}{5pt plus 2pt minus 1pt}
\setlength{\belowdisplayshortskip}{7pt plus 2pt minus 1pt}
\setlength{\jot}{4.5pt}

\section{PICNN-Assisted Physics-Preserving Scheme}

\indent In this section, we construct a PICNN-assisted physics-preserving scheme for the reduced pressure-saturation formulation \eqref{eq:reduced_system}. The discretization scheme follows the mGSAV-LM framework \cite{li2025class}. In this work, the prediction step is performed by the PICNN, whose loss function is built from finite-volume residuals of the reduced system. The predicted fields are then processed by the mGSAV relaxation and the LM-KKT correction. The mGSAV step is used to maintain consistency with the discrete energy law, while the LM-KKT correction enforces the prescribed mass balance and the saturation bounds.

Firstly, we rewrite the model \eqref{eq:model_mass}-\eqref{eq:model_constraint} as the mGSAV-Lagrange reformulation in order to construct the physics-preserving scheme. Let us introduce the following SAV
\begin{flalign}
&\modelindent
Q(t)
=
\mathcal E(t)+\kappa
=
\int_{\Omega}\phi F(S_w)\mathrm{~d}x+\kappa>0, &&
\label{eq:Q_def_picnn}
\end{flalign}
where $\kappa>0$ is a positive constant.
In order to impose the saturation bounds, we introduce the quadratic function
$f(S_w)=(1-S_{rn}-S_w)(S_w-S_{rw})$. Then the admissible interval
$S_{rw}\le S_w\le 1-S_{rn}$ is characterized by $f(S_w)\ge 0$. We further
introduce a LM function $\Phi_w(x,t)$ for the bounds constraint
and a scalar LM $\Psi(t)$ for the mass constraint. In this way,
the following expanded system is obtained:
\begin{subequations}\label{eq:expanded_system_picnn}
\begin{flalign}
&\modelindent
-\nabla\cdot
\left(
\lambda_t(S_w)K\nabla p_w
\right)
+
\nabla\cdot
\left(
\lambda_n(S_w)K\nabla\mu(S_w)
\right)
=
q_t, &&
\label{eq:expanded_pressure_picnn}\\[0.45em]
&\modelindent
\phi\frac{\partial S_w}{\partial t}
-
\nabla\cdot
\left(
\lambda_w(S_w)K\nabla p_w
\right)
=
q_w+\Phi_w f'(S_w)+\Psi, &&
\label{eq:expanded_saturation_picnn}\\[0.45em]
&\modelindent
\Phi_w\ge 0,\qquad f(S_w)\ge 0,\qquad
\Phi_w f(S_w)=0, &&
\label{eq:kkt_picnn}\\[0.45em]
&\modelindent
\left(
\phi\frac{\partial S_w}{\partial t},1
\right)
+
\left\langle
\mathbf{u}_w\cdot \mathbf{n},1
\right\rangle
=
(q_w,1), &&
\label{eq:mass_constraint_picnn}\\[0.45em]
&\modelindent
\frac{\mathrm{d}Q}{\mathrm{d}t}
=
-
\frac{Q(t)}
{\mathcal E(t)+\kappa}
\sum_{\alpha=w,n}
\left(
\left\|
(\lambda_{\alpha}K)^{1/2}\nabla p_{\alpha}
\right\|^2
+
\left\langle
\mathbf{u}_{\alpha}\cdot \mathbf{n},p_{\alpha}
\right\rangle
-
(q_{\alpha},p_{\alpha})
\right). &&
\label{eq:Q_equation_picnn}
\end{flalign}
\end{subequations}
Here $p_n=p_w-\mu(S_w)$, and \eqref{eq:kkt_picnn} gives the KKT
complementarity condition. The multipliers $\Phi_w$ and $\Psi$ are introduced
only for the correction procedure; they enforce the saturation bounds and the
mass constraint after the PICNN prediction.

\subsection{Finite-volume residual training of the PICNN predictor}

\indent This subsection gives the concrete residual construction used in the prediction step. At each time step, the input tensor contains the previous pressure and saturation fields, the normalized permeability and porosity fields, and the two normalized coordinate fields. The coordinate channels store the cell-center locations after rescaling and supply positional information to the convolutional network. The normalization of the medium and coordinate fields is used only for the neural-network input; all coefficients in the finite-volume residuals retain their physical scales.

The network outputs two vectors, denoted by $p_{\rm raw}$ and $S_{\rm raw}$. The pressure approximation is recovered by
\begin{flalign*}
&\modelindent \widetilde p_i=p_{\rm scale}(p_{\rm raw})_i, &&
\end{flalign*}
where $p_{\rm scale}>0$ is a characteristic pressure scale. For closed no-flow problems, it is computed from the explicit part of the chemical potential, for example
\begin{flalign}
&\modelindent p_{\rm scale}
=\frac{1}{N}\sum_{i=1}^{N}
\left|
\sigma_{w,i}\ln S_{w,i}^{*,n}
-\sigma_{n,i}\ln(1-S_{w,i}^{*,n})
+\sigma_{wn,i}
\right|
+\varepsilon_p, &&
        \label{eq:pressure_scale}
\end{flalign}
where $N$ is the number of control volumes, $S_{w,i}^{*,n}$ is the clipped previous-time saturation, and $\varepsilon_p>0$ prevents a zero scale. For injection-driven or open-boundary problems, $p_{\rm scale}$ is chosen from the prescribed pressure or a characteristic pressure difference. The saturation output is used as the intermediate Step-1 approximation,
\begin{flalign*}
&\modelindent \widetilde S_{w,i}=(S_{\rm raw})_i. &&
\end{flalign*}

The wetting-phase, non-wetting-phase, and total mobilities are evaluated explicitly from the saturation at the previous time level:
\begin{flalign}
&\modelindent
\lambda_w(S_w^n)=\frac{k_{rw}(S_w^n)}{\eta_w},\qquad
\lambda_n(S_w^n)=\frac{k_{rn}(S_w^n)}{\eta_n},\qquad
\lambda_t=\lambda_w+\lambda_n. &&
        \label{eq:picnn_mobility_explicit}
\end{flalign}
The chemical potential entering the pressure equation is evaluated in the semi-explicit form
\begin{flalign}
&\modelindent
\widetilde\mu_i
=\sigma_{w,i}\ln S_{w,i}^{*,n}
-\sigma_{n,i}\ln(1-S_{w,i}^{*,n})
+\sigma_{wn,i}(1-2\widetilde S_{w,i}). &&
        \label{eq:picnn_mu_semiexplicit}
\end{flalign}
Here the logarithmic terms are evaluated from the clipped previous-time saturation, while the linear term is evaluated from the Step-1 saturation prediction.

The flux terms are discretized by a cell-centered finite-volume method with two-point flux approximation. For neighboring control volumes $i$ and $j$, set
\begin{flalign*}
&\modelindent
a_{\alpha,i}=K_i\lambda_{\alpha,i},\qquad \alpha\in\{w,n,t\}, &&
\end{flalign*}
and use the harmonic transmissibility
\begin{flalign*}
&\modelindent
T_{\alpha,ij}
=\frac{1}{h^2}
\frac{2a_{\alpha,i}a_{\alpha,j}}{a_{\alpha,i}+a_{\alpha,j}},
\qquad \alpha\in\{w,n,t\}. &&
\end{flalign*}
On no-flow boundaries, the exterior flux is set to zero. Substituting the PICNN output into the finite-volume equations gives the pressure residual
\begin{flalign}
&\modelindent
R_{p,i}
=\sum_{j\in\mathcal N(i)}
T_{t,ij}(\widetilde p_j-\widetilde p_i)
+\sum_{j\in\mathcal N(i)}
T_{n,ij}(\widetilde\mu_j-\widetilde\mu_i), &&
        \label{eq:picnn_pressure_residual}
\end{flalign}
and the saturation residual
\begin{flalign}
&\modelindent
R_{s,i}
=\sum_{j\in\mathcal N(i)}
T_{w,ij}(\widetilde p_j-\widetilde p_i)
+\phi_i\frac{\widetilde S_{w,i}-S_{w,i}^{n}}{\Delta t}. &&
        \label{eq:picnn_saturation_residual}
\end{flalign}
Here $\mathcal N(i)$ denotes the neighboring control volumes of cell $i$. Source terms and boundary fluxes are added to the corresponding residuals for injection-driven or open-boundary problems. For closed systems, the pressure equation is of pure Neumann type; a pressure-gauge constraint is imposed by replacing one pressure residual component with the mean value of $p_{\rm raw}$.

Because the pressure and saturation residuals have different physical scales, row-wise normalization is applied before forming the loss. The pressure and saturation scaling factors are
\begin{flalign}
&\modelindent
c_i^p
=p_{\rm scale}\sum_{j\in\mathcal N(i)} T_{t,ij}
+2|\sigma_{wn,i}|\sum_{j\in\mathcal N(i)} T_{n,ij}, &&
        \label{eq:picnn_pressure_scaling}
\end{flalign}
and
\begin{flalign}
&\modelindent
c_i^s
=p_{\rm scale}\sum_{j\in\mathcal N(i)} T_{w,ij}
+\frac{\phi_i}{\Delta t}. &&
        \label{eq:picnn_saturation_scaling}
\end{flalign}
The scaled residuals are
\begin{flalign}
&\modelindent
\widehat R_{p,i}=\frac{R_{p,i}}{c_i^p+\varepsilon_{\rm sc}},
\qquad
\widehat R_{s,i}=\frac{R_{s,i}}{c_i^s+\varepsilon_{\rm sc}}, &&
        \label{eq:picnn_scaled_residuals}
\end{flalign}
where $\varepsilon_{\rm sc}>0$ avoids division by zero.

The pressure-equation and saturation-equation losses are defined by
\begin{flalign}
&\modelindent
\mathcal L_p
=\operatorname{MSE}(\widehat R_p)
+w_{\rm ms}\mathcal L_{p,{\rm ms}}
+w_{\rm sp}\mathcal L_{p,{\rm sp}}, &&
        \label{eq:picnn_pressure_loss}
\end{flalign}
and
\begin{flalign}
&\modelindent
\mathcal L_s
=\operatorname{MSE}(\widehat R_s)
+w_{\rm ms}\mathcal L_{s,{\rm ms}}
+w_{\rm sp}\mathcal L_{s,{\rm sp}}. &&
        \label{eq:picnn_saturation_loss}
\end{flalign}
The multiscale residual penalty is
\begin{flalign}
&\modelindent
\mathcal L_{\alpha,{\rm ms}}
=\sum_{\ell\in\mathcal B}
\omega_\ell
\operatorname{MSE}\left(B_\ell \widehat R_\alpha\right),
\qquad \alpha\in\{p,s\}, &&
        \label{eq:picnn_multiscale_loss}
\end{flalign}
where $\mathcal B$ is the set of block sizes and $B_\ell$ is the block-averaging operator. The spectral penalty is
\begin{flalign}
&\modelindent
\mathcal L_{\alpha,{\rm sp}}
=\sum_{\boldsymbol{k}\ne\boldsymbol{0}}
\frac{|\widehat R_\alpha(\boldsymbol{k})|^2}{\lambda_{\boldsymbol{k}}},
\qquad \alpha\in\{p,s\}, &&
        \label{eq:picnn_spectral_loss}
\end{flalign}
where $\widehat R_\alpha(\boldsymbol{k})$ is the discrete Fourier coefficient of the scaled residual and $\lambda_{\boldsymbol{k}}$ is the corresponding discrete Laplacian eigenvalue.

At the beginning of each time step, the initial losses $\mathcal L_{p,0}$ and $\mathcal L_{s,0}$ are recorded. The total loss minimized by the optimizer is
\begin{flalign}
&\modelindent
\mathcal J(\theta)
=\frac{\mathcal L_p(\theta)}
{\mathcal L_{p,0}+\varepsilon_{\rm loss}}
+\frac{\mathcal L_s(\theta)}
{\mathcal L_{s,0}+\varepsilon_{\rm loss}}, &&
        \label{eq:picnn_total_loss}
\end{flalign}
where $\varepsilon_{\rm loss}>0$. Transfer learning is used between consecutive time steps: the trained network from the current step initializes the network at the next step, and the output associated with the smallest total loss is taken as the Step-1 prediction.

For problems with source terms or boundary fluxes, the correction step uses the
corresponding time-discrete global mass balance. More precisely, the wetting-phase
target mass is
\begin{flalign}
&\modelindent
M_w^{n+1,{\rm tar}}
=
\sum_{E\in\mathcal T_h}|E|\phi_E S_{w,E}^{n}
+
\Delta t
\left[
(q_w^{n+1},1)
-
\left\langle
\mathbf u_w^{n+1}\cdot\mathbf n,1
\right\rangle
\right]. &&
\label{eq:source_target_mass}
\end{flalign}
The closed no-flow case considered below is the special case
$M_w^{n+1,{\rm tar}}=\sum_{E\in\mathcal T_h}|E|\phi_E S_{w,E}^{n}$.


\subsection{PICNN-assisted physics-preserving scheme for the closed case}

\indent For simplicity,  we consider the closed system with $q_{\alpha}=0$ and $\mathbf{u}_{\alpha}\cdot \mathbf{n}=0$ on $\partial\Omega$. Given $S_w^n$, $p_w^n$ and $Q^n$, the numerical solution at the next time level is obtained by the following four steps.

\textbf{Step 1 (PICNN prediction).} The intermediate  wetting-phase pressure and  saturation are obtained from the residual training procedure described in Section 3.1. At the current time step, the PICNN takes the previous pressure and saturation, the prescribed medium-property channels, and the normalized coordinate channels as inputs and returns
\begin{flalign}
&\modelindent
    \left(
    \widetilde p_w^{\,n+1},
    \widetilde S_w^{\,n+1}
    \right)
    =
    \mathcal N_{\theta}
    \left(
    S_w^n,p_w^n,K,\phi,\boldsymbol{x},\boldsymbol{y}
    \right),
    &&
    \label{eq:picnn_prediction_caseI}
\end{flalign}
where $\mathcal N_{\theta}$ denotes the trained CNN, and $\boldsymbol{x}$ and $\boldsymbol{y}$ denote the normalized coordinate fields. The network parameters are updated by minimizing the finite-volume residual loss \eqref{eq:picnn_total_loss}, with the pressure and saturation residuals defined in \eqref{eq:picnn_pressure_residual} and \eqref{eq:picnn_saturation_residual}, the row-wise scaling in \eqref{eq:picnn_pressure_scaling}, \eqref{eq:picnn_saturation_scaling}, and \eqref{eq:picnn_scaled_residuals}, and the multiscale and spectral penalties in \eqref{eq:picnn_multiscale_loss} and \eqref{eq:picnn_spectral_loss}. Thus, Step 1 only provides the intermediate fields needed by the subsequent mGSAV-LM correction.

For consistency with the semi-explicit pressure residual used during training, the predicted chemical potential is evaluated by
\begin{flalign}
&\modelindent
    \widetilde\mu^{\,n+1}
    =
    \sigma_w\ln(S_w^n)
    -
    \sigma_n\ln(1-S_w^n)
    +
    \sigma_{wn}
    \left(
    1-2\widetilde S_w^{\,n+1}
    \right). &&
    \label{eq:mu_tilde_caseI}
\end{flalign}
Here the logarithmic part is evaluated at the previous saturation, while the linear interfacial contribution is evaluated with the predicted saturation. In practical implementation, the previous saturation is clipped as in \eqref{eq:pressure_scale} and \eqref{eq:picnn_mu_semiexplicit} when the logarithmic terms are evaluated. The predicted non-wetting-phase pressure is then obtained from the reduced pressure relation:
\begin{flalign}
&\modelindent
    \widetilde p_n^{\,n+1}
    =
    \widetilde p_w^{\,n+1}
    -
    \widetilde\mu^{\,n+1}. &&
    \label{eq:pn_tilde_caseI}
\end{flalign}
The exact saturation bounds are not imposed on the raw network output. They are enforced below by the LM-KKT correction, while the energy relaxation uses a clipped version of $\widetilde S_w^{\,n+1}$ only for evaluating the free energy.


\textbf{Step 2 (energy stability).} Given $Q^n$, $\widetilde p_w^{\,n+1}$, and $\widetilde S_w^{\,n+1}$, 
we compute the auxiliary variable $\widetilde Q^{\,n+1}$ by
\begin{flalign}
&\modelindent
\begin{aligned}
\frac{
\widetilde Q^{\,n+1}-Q^n
}
{\Delta t}
={}&
-
\frac{
\widetilde Q^{\,n+1}
}
{
\mathcal E
\left(
\widetilde S_w^{*,n+1}
\right)
+\kappa
}
\sum_{\alpha=w,n}
\left\|
\left(
\lambda_{\alpha}
\left(
\widetilde S_w^{*,n+1}
\right)K
\right)^{1/2}
\nabla_h
\widetilde p_{\alpha}^{\,n+1}
\right\|_h^2,
\end{aligned} &&
\label{eq:Q_tilde_caseI}
\end{flalign}
where
$ \displaystyle
\mathcal E
\left(
\widetilde S_w^{*,n+1}
\right)
=
\int_{\Omega}
\phi F
\left(
\widetilde S_w^{*,n+1}
\right)
\mathrm{~d}x$ 
and $\widetilde S_w^{*,n+1}
= \min
\left\{
\max \left\{ \widetilde S_w^{\,n+1},S_{rw}
\right\}, 1-S_{rn} \right\}$ by using the cut-off technique.
Then, we set
\begin{flalign}
&\modelindent
\xi^{n+1}
=
\frac{
\widetilde Q^{\,n+1}
}
{
\mathcal E
\left(
\widetilde S_w^{*,n+1}
\right)
+\kappa
},
\qquad
\eta^{n+1}
=
1-\left(1-\xi^{n+1}\right)^2, &&
\label{eq:xi_eta_caseI}
\end{flalign}
therefore, we can obtain the energy-corrected saturation $\widehat S_w^{\,n+1}$ by
\begin{flalign}
&\modelindent
\widehat S_w^{\,n+1}
=
\eta^{n+1}
\widetilde S_w^{*,n+1}. &&
\label{eq:S_hat_caseI}
\end{flalign}

\textbf{Step 3 (correction).} Given $\widehat S_w^{\,n+1}$, we compute $\Phi_w^{n+1}(x)$, $\Psi^{n+1}$, and  $S_w^{n+1}$ by
\begin{flalign}
&\modelindent
\phi
\frac{
S_w^{n+1}-\widehat S_w^{\,n+1}
}
{\Delta t}
=
\Phi_w^{n+1}f'
\left(
S_w^{n+1}
\right)
+
\Psi^{n+1}, &&
\label{eq:correction_caseI} \\
&\modelindent
\Phi_w^{n+1}\ge 0,
\qquad
f(S_w^{n+1})\ge 0,
\qquad
\Phi_w^{n+1}f(S_w^{n+1})=0. &&
\label{eq:kkt_caseI} \\
&\modelindent
\left(
\phi
\frac{
S_w^{n+1}-S_w^n
}
{\Delta t},
1
\right)
=
0. &&
\label{eq:mass_caseI}
\end{flalign}
Finally, update the non-wetting phase by
\begin{flalign}
&\modelindent
S_n^{n+1}
=
1-S_w^{n+1}. &&
\label{eq:Sn_caseI}
\end{flalign}

\textbf{Step 4 (mGSAV relaxation).}
For the closed system, the SAV is updated by
\begin{flalign}
&\modelindent
Q^{n+1}
=
\min
\left\{
Q^n,
\mathcal E
\left(
S_w^{n+1}
\right)
+\kappa
\right\}. &&
\label{eq:Q_update_caseI}
\end{flalign}

In what follows, let us describe how to efficient implement of the correction step for the proposed scheme. The LM-KKT correction can be reduced to a scalar nonlinear equation for the mass multiplier $\Psi^{n+1}$. Let $\mathcal T_h$ denote the set of control volumes. For a fixed $\Psi^{n+1}$, define the cell-wise quantity
\begin{flalign}
&\modelindent
\Pi_E^{n+1}
=
\widehat S_{w,E}^{\,n+1}
+
\frac{\Delta t}{\phi_E}\Psi^{n+1},
\qquad E\in\mathcal T_h . &&
\label{eq:Pi_cell}
\end{flalign}
Then the corrected saturation is obtained by the projection
\begin{flalign}
&\modelindent
S_{w,E}^{n+1}(\Psi^{n+1})
=
\min
\left\{
\max
\left\{
\Pi_E^{n+1},S_{rw}
\right\},
1-S_{rn}
\right\},
\qquad E\in\mathcal T_h . &&
\label{eq:projection_cell}
\end{flalign}
This projection is the cell-wise realization of the KKT bounds condition. When $\Pi_E^{n+1}$ lies inside the admissible interval, the local multiplier $\Phi_{w,E}^{n+1}$ is inactive. Otherwise, the projection fixes the saturation at the corresponding bound.

The scalar multiplier $\Psi^{n+1}$ is determined by substituting \eqref{eq:projection_cell} into the mass constraint. The scalar equation is
\begin{flalign}
&\modelindent
N(\Psi^{n+1})
=
\sum_{E\in\mathcal T_h}
|E|\phi_E
S_{w,E}^{n+1}(\Psi^{n+1})
-
\sum_{E\in\mathcal T_h}
|E|\phi_E
S_{w,E}^{n}
=
0 . &&
\label{eq:N_caseI}
\end{flalign}
For a problem with source terms or nonzero boundary fluxes, the second term in
\eqref{eq:N_caseI} is replaced by the target mass
$M_w^{n+1,{\rm tar}}$ in \eqref{eq:source_target_mass}.
The scalar nonlinear equation \eqref{eq:N_caseI} can be solved efficiently by the secant method. Once $\Psi^{n+1}$ is obtained, the corrected saturation is computed from \eqref{eq:projection_cell}. Hence,
\begin{flalign}
&\modelindent
S_{rw}
\le
S_{w,E}^{n+1}
\le
1-S_{rn},
\qquad E\in\mathcal T_h . &&
\label{eq:bounds_w_final}
\end{flalign}
Using $S_n^{n+1}=1-S_w^{n+1}$, we also have
\begin{flalign}
&\modelindent
S_{rn}
\le
S_{n,E}^{n+1}
\le
1-S_{rw},
\qquad E\in\mathcal T_h . &&
\label{eq:bounds_n_final}
\end{flalign}

In summary, the proposed scheme follows the correction structure of the mGSAV-LM method \cite{li2025class}, while the prediction step is performed by a PICNN constrained by conservative finite-volume residuals. The final accepted solution is obtained after the energy correction and the LM-KKT correction, rather than being taken directly from the neural-network output.

\subsection{Structure-preserving properties}

The correction part of the proposed method is inherited from the mGSAV-LM framework \cite{li2025class}. Hence the PICNN output is used only as an intermediate prediction, while the final accepted solution satisfies the same discrete physical constraints after the correction steps. We summarize these properties in the following theorem.

\refstepcounter{thm}\label{thm:structure_preserving}
\noindent{\bf Theorem \thethm.}
{\it Assume that the closed system satisfies $q_\alpha=0$ and
$\mathbf u_\alpha\cdot\mathbf n=0$ on $\partial\Omega$, and that
$S_w^n$ lies in the admissible interval. Let $Q^n>0$ and choose $\kappa$
such that $\mathcal E(S)+\kappa>0$ for all admissible saturations. If the
correction equation \eqref{eq:N_caseI} is solved, then the solution produced
by Steps 1--4 satisfies the following properties:
\begin{enumerate}[(i)]
    \item Phase-wise mass conservation.
    \begin{flalign}
    &\modelindent
    \sum_{E\in\mathcal T_h}|E|\phi_E S_{w,E}^{n+1}
    =
    \sum_{E\in\mathcal T_h}|E|\phi_E S_{w,E}^{n},
    \qquad
    \sum_{E\in\mathcal T_h}|E|\phi_E S_{n,E}^{n+1}
    =
    \sum_{E\in\mathcal T_h}|E|\phi_E S_{n,E}^{n}. &&
    \label{eq:theorem_mass}
    \end{flalign}
    \item Modified energy dissipation.
    \begin{flalign}
    &\modelindent
    Q^{n+1}\le Q^n,
    \qquad
    Q^{n+1}\le \mathcal E(S_w^{n+1})+\kappa . &&
    \label{eq:theorem_energy}
    \end{flalign}
    \item Saturation bounds preservation.
    \begin{flalign}
    &\modelindent
    S_{rw}\le S_{w,E}^{n+1}\le 1-S_{rn},
    \qquad
    S_{rn}\le S_{n,E}^{n+1}\le 1-S_{rw},
    \qquad E\in\mathcal T_h . &&
    \label{eq:theorem_bounds}
    \end{flalign}
\end{enumerate}
}

\noindent{\bf Proof.}
The wetting-phase mass identity follows directly from the scalar constraint
\eqref{eq:mass_caseI}, or equivalently from \eqref{eq:N_caseI}. Since
$S_n^{n+1}=1-S_w^{n+1}$ and the mesh and porosity are fixed, the
non-wetting-phase mass is conserved as well.
The modified energy dissipation follows because Step 4 sets
$Q^{n+1}$ to the minimum of $Q^n$ and
$\mathcal E(S_w^{n+1})+\kappa$ in \eqref{eq:Q_update_caseI}. Hence both
inequalities in \eqref{eq:theorem_energy} hold. Finally, the projection formula
\eqref{eq:projection_cell} gives the bounds for $S_w^{n+1}$, and the bounds
for $S_n^{n+1}$ follow from $S_n^{n+1}=1-S_w^{n+1}$. This is the same
correction argument as in the physics-preserving mGSAV-LM scheme
\cite{li2025class}. \hfill$\square$

\raggedbottom
\section{Numerical Experiments}

In this section, several numerical experiments are presented to examine the performance of the proposed PICNN-assisted physics-preserving scheme. The spatial discretization is carried out on a cell-centered finite-volume grid. The PICNN is used to generate the intermediate pressure and saturation fields, and the mGSAV-LM correction is then applied to obtain the final corrected solution.

The phase mobilities are defined by \eqref{eq:picnn_mobility_explicit}, and the relative permeabilities are taken in the standard power-law form
$k_{rw}(S_w)=S_w^{\iota}$ and $k_{rn}(S_n)=S_n^{\iota}$ in all computations. The value of the exponent $\iota$, the viscosities, the residual saturations, and the energy parameters $\sigma_w$, $\sigma_n$, and $\sigma_{wn}$ are specified in each example.
In the mGSAV correction, the relaxation factor is computed by the relation in \eqref{eq:xi_eta_caseI}. The scalar nonlinear equation for the LM $\Psi^{n+1}$ is reduced to \eqref{eq:N_caseI} and is solved by the secant method. The iteration is stopped once the residual of the scalar equation is below the prescribed tolerance $\epsilon_\Psi$. In the following computations, we take $\epsilon_{\Psi}=10^{-10}$ and set the maximum number of secant iterations to $100$, unless otherwise stated.

In Example 1, the manufactured source term is included in the target mass
\eqref{eq:source_target_mass}; hence the reported mass errors are measured
against the source-updated phase masses. In Example 2, there are no source terms
and no boundary fluxes, so the correction reduces to the closed mass constraint
\eqref{eq:N_caseI}.

For the PICNN predictor, the main training settings, including the number of epochs, learning rate, and selected loss weights, are reported in the corresponding examples when needed.

\subsection{Example 1}

\indent We first consider a manufactured-solution test for the two-phase flow model. The source terms $q_t$ and $q_w$ are chosen so that the problem admits the following manufactured wetting-phase saturation and pressure:
\begin{flalign*}
&\modelindent
\left\{
\begin{aligned}
S_w(x,y,t)
&=
e^{-t}\left(\frac{1}{16}\cos(\pi x)\cos(\pi y)+0.52\right),\\
p_w(x,y,t)
&=
0.55e^{-t}\cos(\pi x)\cos(\pi y),\\
S_{rw}&=0.1,\\
S_{rn}&=0.1.
\end{aligned}
\right. &&
\end{flalign*}
The initial condition is obtained by evaluating the manufactured solution at $t=0$. The numerical, physical, and PICNN parameters used in this example are listed in Table \ref{tab:case1_parameters}.

\begin{table}[H]
    \centering
    \caption{Numerical, physical, and PICNN parameters for Example 1.}
    \label{tab:case1_parameters}
    \small
    \renewcommand{\arraystretch}{1.13}
    \begin{tabular}{lll}
        \toprule
        Quantity & Symbol & Value \\
        \midrule
        Computational domain & $\Omega$ & $[0,1]\times[0,1]$ \\
        Grid size & $N_x\times N_y$ & $16\times16$ \\
        Mesh size & $h$ & $0.0625$ \\
        Time step & $\Delta t$ & $0.00390625$ \\
        Final time & $T$ & $1$ \\
        Reference saturation & $S_{w0}$ & $0.52$ \\
        Porosity & $\phi$ & $0.95$ \\
        Permeability & $K$ & $1.1\times10^{-4}$ \\
        Residual saturations & $S_{rw},S_{rn}$ & $0.1,\ 0.1$ \\
        Mobility parameters & $\eta_w,\eta_n,\iota$ & $1.05,\ 0.55,\ 3$ \\
        Model parameters & $\sigma_w,\sigma_n,\sigma_{wn}$ & $0.60,\ 0.055,\ 0.34$ \\
        Training epochs & first/later steps & $500/100$ \\
        \bottomrule
    \end{tabular}
\end{table}

\begin{figure}[H]
    \centering
    \includegraphics[width=0.86\textwidth]{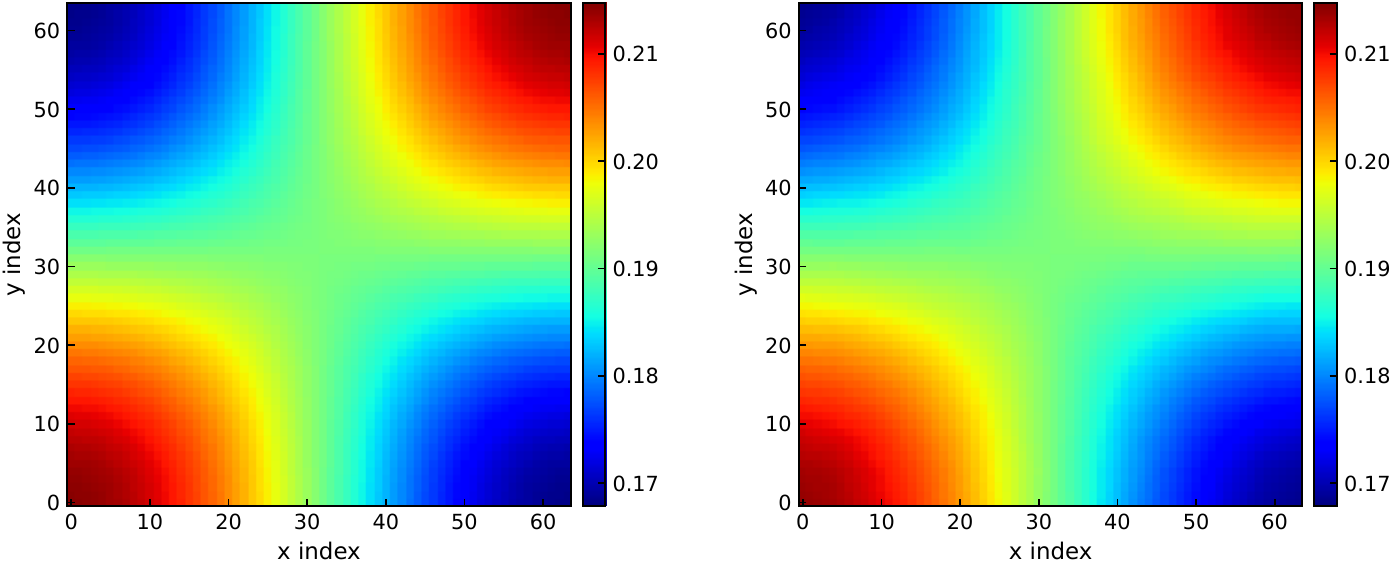}
    \caption{Exact and numerical solutions of wetting-phase saturation at $T=1$ (Example 1).}
    \label{fig:case1_exact_num}
\end{figure}

Figure \ref{fig:case1_exact_num} compares the exact and numerical wetting-phase saturations at the final time. The numerical result reproduces the main spatial structure of the manufactured solution: the locations of the high- and low-saturation regions are correctly captured, and the overall amplitude agrees well with the exact field. The corrected solution also remains strictly inside the prescribed saturation interval, which indicates that the post-processing steps do not destroy the principal spatial profile of the PICNN prediction.

\begin{figure}[H]
    \centering
    \includegraphics[width=0.80\textwidth]{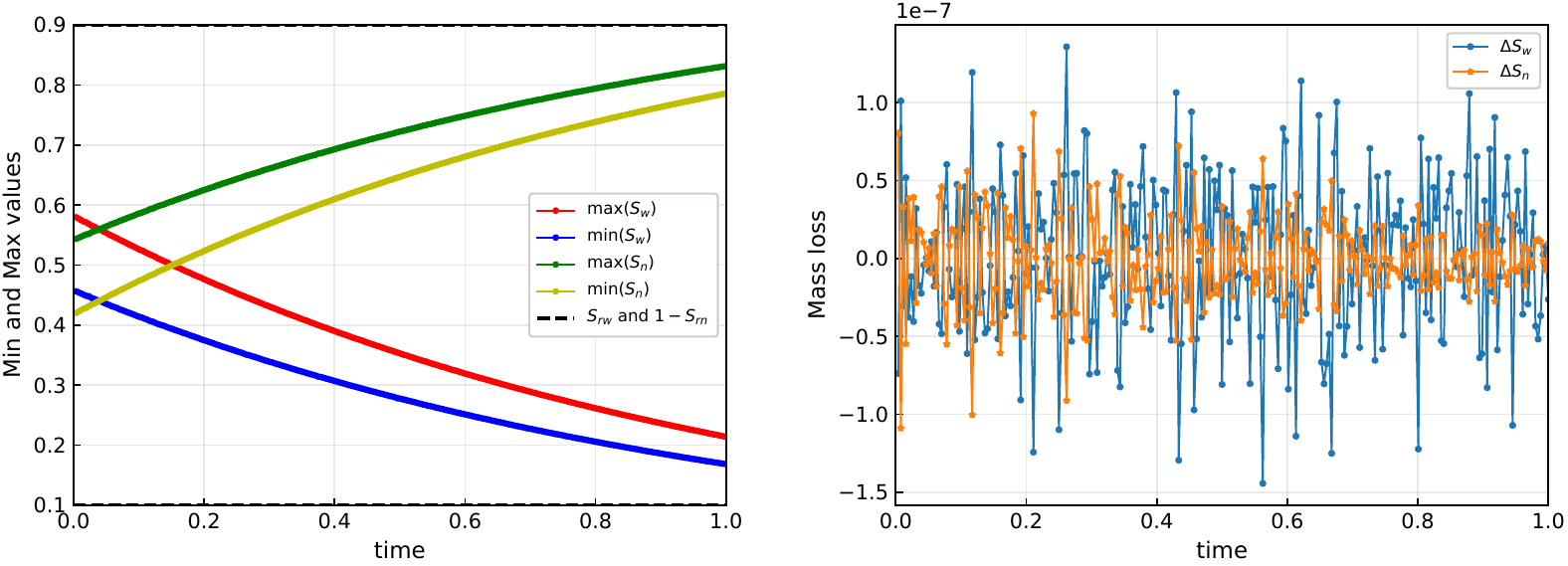}
    \caption{Boundness of two phase saturations and phase-wise relative mass errors $T=1$ (Example 1).}
    \label{fig:case1_bounds_mass}
\end{figure}
The bounds-preserving and mass-conservation diagnostics are shown in Figure \ref{fig:case1_bounds_mass}. The extrema of both phase saturations remain within the admissible range $[0.1,0.9]$ during the whole simulation. The relative mass errors of the wetting and non-wetting phases oscillate around zero and remain at approximately the $10^{-7}$ level, demonstrating that the correction procedure effectively controls the accumulated mass drift caused by the approximate PICNN prediction.

\begin{figure}[H]
    \centering
    \includegraphics[width=0.38\textwidth]{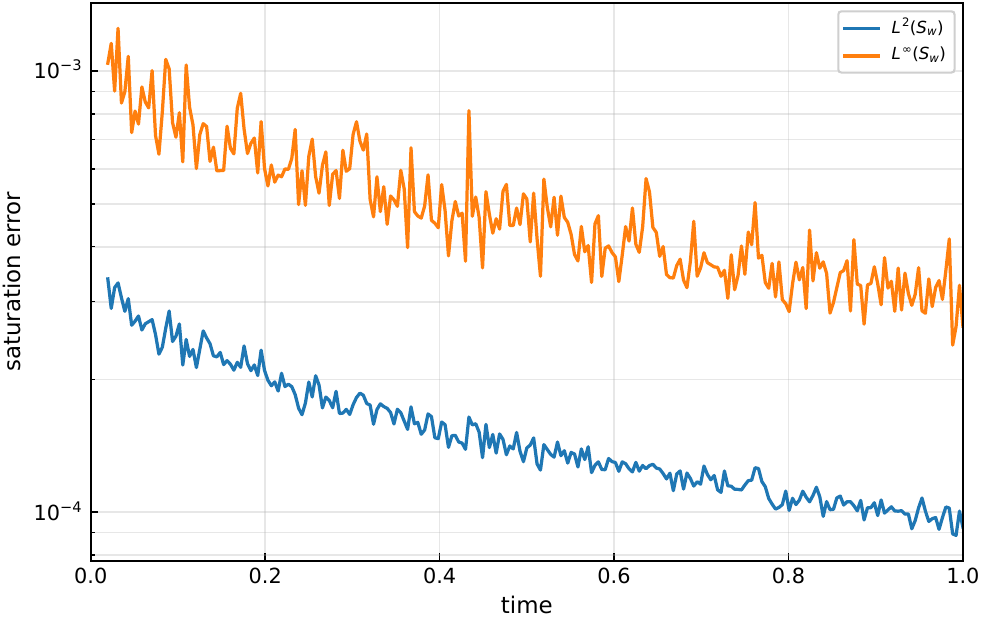}
    \caption{Temporal evolution of the $L^2$ and $L^\infty$ errors of $S_w$ (Example 1).}
    \label{fig:case1_saturation_error}
\end{figure}

Figure \ref{fig:case1_saturation_error} gives the temporal evolution of the saturation errors. Both the $L^2$ and $L^\infty$ errors remain small over the full time interval, and the final diagnostic values are summarized in Table \ref{tab:case1_diagnostics}.

\begin{table}[htbp]
    \centering
    \caption{Final diagnostic quantities (Example 1).}
    \label{tab:case1_diagnostics}
    \footnotesize
    \renewcommand{\arraystretch}{1.0}
    \begin{tabular}{lc}
        \toprule
        Diagnostic quantity & Value \\
        \midrule
        $\|S_w^{\rm num}-S_w^{\rm ex}\|_{L^2}$ & $9.2479\times10^{-5}$ \\
        $\|S_w^{\rm num}-S_w^{\rm ex}\|_{L^\infty}$ & $2.6435\times10^{-4}$ \\
        Wetting-phase relative mass error & $-2.6264\times10^{-8}$ \\
        Non-wetting-phase relative mass error & $9.8494\times10^{-9}$ \\
        \bottomrule
    \end{tabular}
\end{table}

Overall, this manufactured-solution test shows that the PICNN predictor combined with Steps 2--4 gives a reasonable approximation of the wetting-phase saturation. The numerical solution is close to the exact solution, the saturation remains within the prescribed interval, and the phase mass errors stay at a small level. These results indicate that the post-processing steps do not destroy the accuracy of the PICNN prediction and can enforce the desired bound-preserving and mass-conservation properties in this test.

\subsection{Example 2}

\indent We next consider two-phase flow in a closed heterogeneous square domain $\Omega=[0,10]\times[0,10]\ {\rm m}^2$. No-flow boundary conditions are imposed on the whole boundary, and no injection or production terms are included. The initial wetting-phase saturation is set to $S_w^0=0.4$, with $S_n^0=1-S_w^0$. The heterogeneous permeability and porosity fields are specified by the two-region parameters listed in Table~\ref{tab:case3_physical}.

\begin{table}[htbp]
    \centering
    \caption{Physical parameters (Example 2).}
    \label{tab:case3_physical}
    \small
    \renewcommand{\arraystretch}{1.15}
    \begin{tabular}{lccccc}
        \toprule
        Region & $K$ (md) & $\phi$ & $\sigma_w$ (bar) & $\sigma_n$ (bar) & $\sigma_{wn}$ (bar) \\
        \midrule
        High-permeability region & $25$ & $0.30$ & $1.1677$ & $0.1007$ & $0.7248$ \\
        Low-permeability region & $15$ & $0.20$ & $1.5074$ & $0.1300$ & $0.9357$ \\
        \bottomrule
    \end{tabular}
\end{table}

\begin{table}[htbp]
\centering
\caption{Numerical and PICNN parameters (Example 2).}
\label{tab:case3_numerical}
\begin{tabular}{lll}
\hline
Parameter & Value & Description \\
\hline
Grid size & $50\times 50$ & cell-centered grid \\
Mesh size & $h=0.2\,{\rm m}$ & uniform mesh \\
Time step & $\Delta t=0.9\,{\rm s}$ & first-order time stepping \\
Final time & $T=10\,{\rm h}$ & $N_t=40000$ steps \\
Residual saturations & $S_{rw}=S_{rn}=10^{-3}$ & admissible interval \\
Viscosities & $\eta_w=1\,{\rm cP},\ \eta_n=0.5\,{\rm cP}$ & phase viscosities \\
Learning rate & $10^{-3}/10^{-4}$ & early/later steps \\
Training epochs & $2000,\ 100$--$200$ & first step/later steps \\
\hline
\end{tabular}
\end{table}

\begin{figure}[H]
    \centering
    \includegraphics[width=0.84\textwidth]{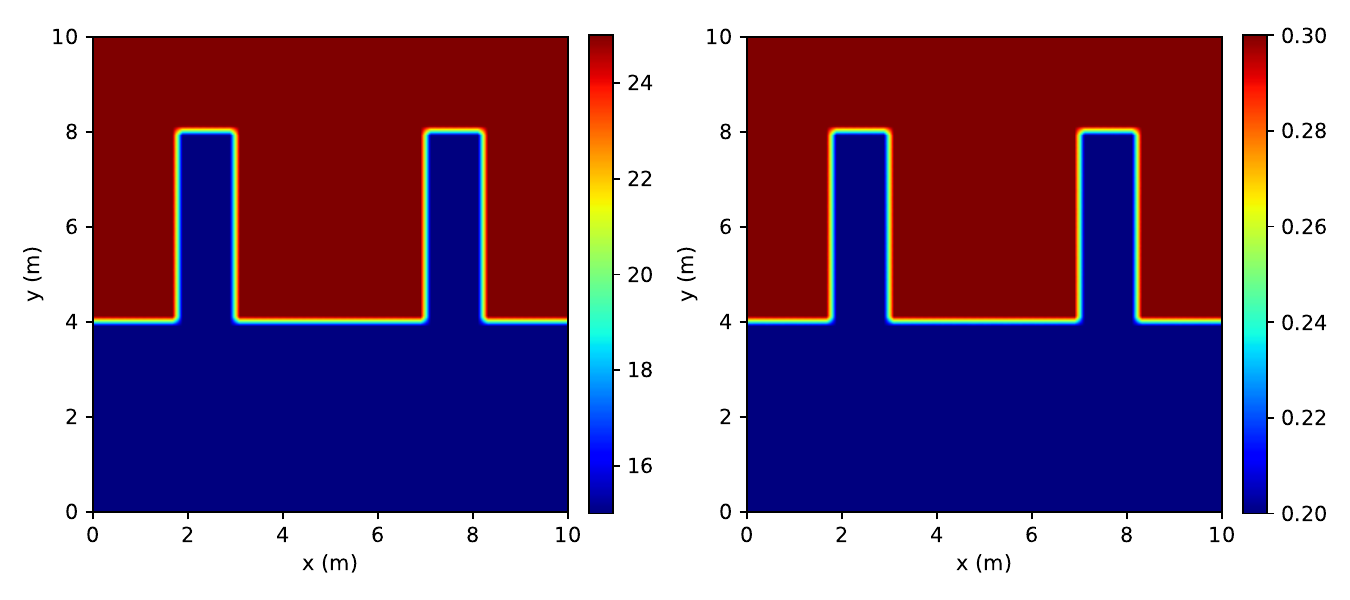}
    \caption{Distributions of permeability (left) and porosity (right) with mesh size $h=0.2\,{\rm m}$ (Example 2).}
    \label{fig:case3_medium}
\end{figure}

Figure \ref{fig:case3_medium} shows the heterogeneous permeability and porosity fields with mesh size $h=0.2\,{\rm m}$. The low-permeability and low-porosity region consists of a lower block connected with two vertical channels, while the remaining part of the domain is occupied by the high-permeability region. This geometry introduces sharp material interfaces and therefore provides a meaningful test for the stability of the harmonic-average flux discretization and for the robustness of the PICNN-assisted correction procedure.

\begin{figure}[H]
    \centering
    \includegraphics[width=0.285\textwidth,trim=0pt 151.2pt 316.8pt 0pt,clip]{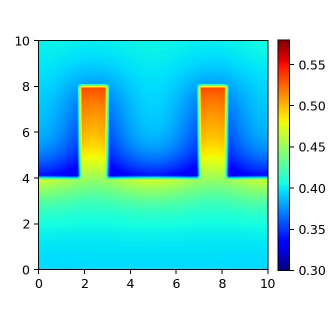}
    \hspace{0.028\textwidth}
    \includegraphics[width=0.285\textwidth,trim=158.4pt 151.2pt 158.4pt 0pt,clip]{test2_figures_pdf/fig5_4_wetting_saturation.pdf}
    \hspace{0.028\textwidth}
    \includegraphics[width=0.285\textwidth,trim=316.8pt 151.2pt 0pt 0pt,clip]{test2_figures_pdf/fig5_4_wetting_saturation.pdf}\\[0.75em]
    \includegraphics[width=0.285\textwidth,trim=0pt 0pt 316.8pt 151.2pt,clip]{test2_figures_pdf/fig5_4_wetting_saturation.pdf}
    \hspace{0.028\textwidth}
    \includegraphics[width=0.285\textwidth,trim=158.4pt 0pt 158.4pt 151.2pt,clip]{test2_figures_pdf/fig5_4_wetting_saturation.pdf}
    \hspace{0.028\textwidth}
    \includegraphics[width=0.285\textwidth,trim=316.8pt 0pt 0pt 151.2pt,clip]{test2_figures_pdf/fig5_4_wetting_saturation.pdf}
    \caption{Distributions of wetting-phase saturation with mesh size $h=0.2\,{\rm m}$ for the present PICNN-assisted scheme: from left to right and top to bottom, $0.25\,{\rm h}$, $1.25\,{\rm h}$, $2.5\,{\rm h}$, $3.75\,{\rm h}$, $6.25\,{\rm h}$, and $10\,{\rm h}$ (Example 2).}
    \label{fig:case3_saturation}
\end{figure}

Since $S_w+S_n=1$, only the wetting-phase saturation is displayed in Figure \ref{fig:case3_saturation}. The figure presents the distributions of wetting-phase saturation at $0.25\,{\rm h}$, $1.25\,{\rm h}$, $2.5\,{\rm h}$, $3.75\,{\rm h}$, $6.25\,{\rm h}$, and $10\,{\rm h}$. In the absence of external injection or production, the saturation redistribution is mainly driven by the chemical-potential gradient induced by the heterogeneous medium. At early times, the wetting phase rapidly accumulates in the low-permeability channels, while the saturation in the high-permeability part decreases. As time advances, the spatial gradients gradually weaken; by the interval from $6.25\,{\rm h}$ to $10\,{\rm h}$, the solution is already close to a steady distribution.

\begin{figure}[H]
    \centering
    \includegraphics[width=0.54\textwidth]{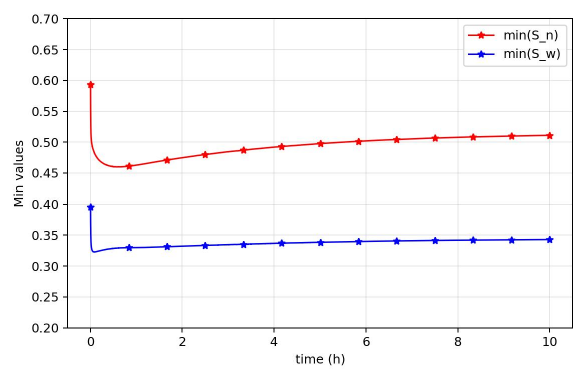}
    \caption{Minimum values of two phase saturations (Example 2).}
    \label{fig:case3_bounds}
\end{figure}

The bounds-preserving property is reported in Figure \ref{fig:case3_bounds}. During the computation, the minimum value of $S_w$ is approximately $0.3228121$, and the minimum value of $S_n$ is approximately $0.45991558$, both of which are far above the residual saturation $10^{-3}$. At the final time, $\min S_w=0.34294483$ and $\min S_n=0.51110297$. Hence, the combination of the PICNN prediction and the correction steps maintains the physical saturation range over all $40000$ time steps, without producing negative saturations or upper-bound violations.

\begin{figure}[H]
    \centering
    \includegraphics[width=0.88\textwidth]{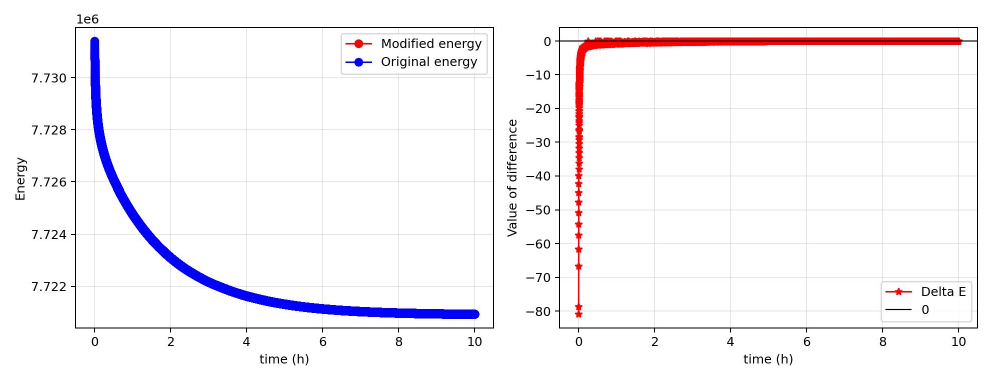}
    \caption{Modified energy, original energy, and energy increments for Example 2.}
    \label{fig:case3_energy}
\end{figure}

The energy behavior is shown in Figure \ref{fig:case3_energy}. The modified energy decreases monotonically from $7.731413\times10^6$ to $7.720931\times10^6$, which is consistent with the expected discrete dissipation mechanism. We define
\begin{flalign}
&\modelindent
\Delta E^n=\mathcal E(S_w^{n+1})+\kappa-Q^n . &&
\label{eq:case3_energy_increment}
\end{flalign}
The increment plot indicates that $\Delta E^n$ remains close to zero and is dominated by nonpositive values. A few tiny positive deviations appear because Step 1 is solved by a finite number of neural-network optimization iterations. The largest positive value is about $3.55\times10^{-2}$, whose relative size with respect to the energy scale $7.7\times10^6$ is about $4.6\times10^{-9}$. The corrected energy sequence itself is nonincreasing over all time steps.

To evaluate mass conservation, we define the relative mass error for phase $\alpha\in\{w,n\}$ by
\begin{flalign}
&\modelindent
\Delta S_\alpha^k
=
\frac{
\sum_{E\in\mathcal T_h}h^2\phi_E S_{\alpha,E}^k
-
\sum_{E\in\mathcal T_h}h^2\phi_E S_{\alpha,E}^0
}{
\sum_{E\in\mathcal T_h}h^2\phi_E S_{\alpha,E}^0
}.
&&
\label{eq:case3_mass_error}
\end{flalign}

\begin{figure}[H]
    \centering
    \includegraphics[width=0.88\textwidth]{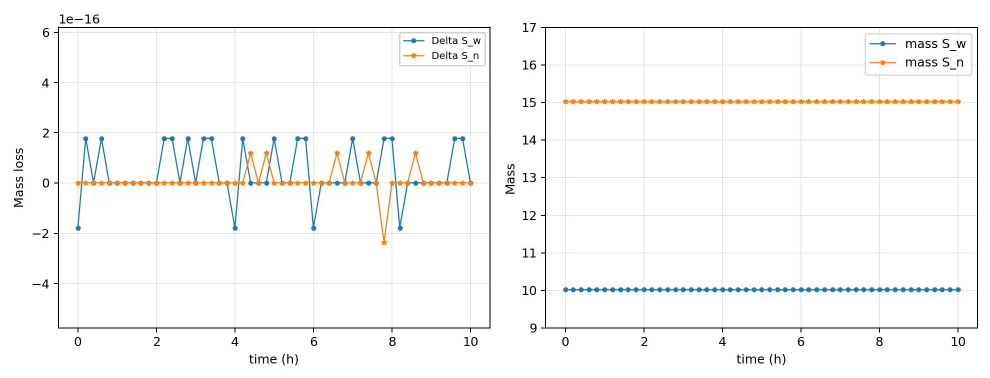}
    \caption{Relative phase-mass errors and discrete phase masses for Example 2.}
    \label{fig:case3_mass}
\end{figure}

Figure \ref{fig:case3_mass} shows that the relative mass errors of both phases stay at the level of $10^{-16}$. More precisely, the wetting-phase error lies approximately in $[-5.3206,3.5470]\times10^{-16}$, and the non-wetting-phase error lies approximately in $[-3.5470,3.5470]\times10^{-16}$. The discrete wetting- and non-wetting-phase masses remain fixed at $10.016$ and $15.024$, respectively. These results confirm that the global mass correction eliminates the cumulative mass drift from the approximate PICNN solve and preserves phase mass to machine precision in the long-time simulation.

In summary, together with Example 1, this test shows that the proposed PICNN-assisted framework stably resolves saturation redistribution under heterogeneous material parameters while preserving the admissible bounds, energy-dissipation behavior, and global phase mass.

\section{Conclusion}

This paper proposed a PICNN-assisted physics-preserving framework for the thermodynamically consistent model of incompressible and immiscible two-phase flow in porous media. The PICNN predictor provides intermediate pressure and saturation fields, while the subsequent mGSAV-LM correction is used to preserve the main physical structures of the model, including energy dissipation, phase-mass conservation, and saturation bounds. The main advantage of the proposed method is that it combines the local representation ability of CNNs with finite-volume residual training. The CNN captures local spatial interactions among neighboring control volumes in the pressure and saturation fields, while the finite-volume/TPFA residuals incorporate heterogeneous permeability effects through interfacial fluxes. Thus, the predictor learns flow patterns constrained by the discrete two-phase flow equations, instead of relying on pointwise continuous PDE residuals. Moreover, the correction stage separates physical-structure enforcement from the neural-network approximation: even when the PICNN prediction is only approximate, the final solution can still satisfy the desired mass, energy, and bound constraints. This makes the framework attractive for repeated time-marching simulations, where transfer learning can reduce the training effort while the correction step maintains physical reliability.

The numerical examples show that the corrected solutions remain accurate in the manufactured-solution test and stable in the closed heterogeneous porous-medium test. In particular, the correction steps keep the saturation within the admissible interval, maintain phase mass at a very small error level, and preserve the expected energy behavior. These results indicate that combining neural-network prediction with structure-preserving post-processing is a feasible strategy for two-phase flow simulation. Future work may focus on higher-dimensional problems, more complex boundary conditions, and systematic comparisons with traditional solvers.

\bibliographystyle{siam}
\bibliography{references/ref}

\end{document}